# A tool for implementation of a domain model based on fuzzy relationships

Ali AAJLI
Laboratory of Computer Systems and Vision -LabSIV, Faculty of science, Ibn Zohr University
Agadir, Morocco

Karim AFDEL
Laboratory of Computer Systems and Vision LabSIV, Faculty of science, Ibn Zohr University
Agadir, Morocco

*Abstract*— The domain model is one of the important components used by adaptive learning systems to automatically generate customized courses for the learners. In this paper our contribution is to propose a new tool for implementation of a domain model based on fuzzy relationships among concepts. This tool allows the experts and teachers to find the best parameters in order to adapt the learners' differences.

## I. INTRODUCTION

The main aim of adaptive learning systems (ALS) is to customize the educational logic developed in their courses; those systems are considered adaptive when they can dynamically change to better suit the learning in response to information collected during the course of learning such as a learner's profile, or achievement test score.

Many adaptive learning systems for education have emerged and have even influenced a number of recent systems (Brusilovsky, 2000). We cite for example APeLS (Conlan et al., 2005), Saxon Phonics system (Duffy, 2005), INTERBOOK system (Brusilovsky, 1998), the system WELSA (Popescu, 2008), AHA system (De Bra, 1998a), the SmexWeb system (Koch, 2000), the system ELM-ART (Weber et al., 2001), the KOD system (Sampson, 2002), the system calls (Colan 2002), the system Alfanet (Alfanet, 2005), the system Medyna (Behaz, 2008), and the OrPAF system (Yessad, 2009), GRAPPLE project, etc.

Architecture of adaptive learning systems is decomposed into three main models that are required to create or automatically generate customized courses to a learner (De Bra, 2001; Brusilovsky, 2003):
- Domain Model (DM): represents the concepts as hierarchy of learning and relationships between the concepts
- Learner Model (LS): describe the learner's profile such as his knowledge, his characteristics and his preference.
- Adaptation Model (AM): defines the concept selection rules that are used for selecting appropriate concepts from the domain model, as well as, the content selection rules from the media space.

In this paper, we focus on designing domain model based on fuzzy logic approach.

The following section presents an overview of some domain model that used fuzzy logic approach.

## II. OVERVIEW OF SOME EXISTING DOMAIN MODEL BASED ON FUZZY LOGIC APPROACH

Several learning systems build their domain model by using a number of different methods of fuzzy logic (Al-Sarem et al, 2010 and Chu et al., 2010 and Chen and Bai, 2008). Sue et al., 2010, used a two-phase method that extracts the association rules between the skills by applying fuzzy logic to convert the grades learners into three levels of difficulty and construct a learning hierarchy. Bai and Chen, 2010, simplified and improved the latter method in adaptive way.

However, they don't take into account the possibility of using a learning hierarchy predefined by experts of a specific field, and those domain model considered grades obtained by learners during the process learning is a fuzzy notion.

In our previous works (Aajli and Afdel, 2014), we proposed an approach of domain model based on fuzzy logic. The idea behind it combining the hierarchy learning predefined by one expert of a specific field and that developed automatically using the fuzzy logic (Fuzzy Sets Theory), and we consider the relationships of prerequisites between the concepts in the concept map are not definitive and they are fuzzy relationships.

In this paper, we extend this idea by presenting it as a tool implements our domain model based on fuzzy logic and try to answer the following question: what are the best parameters of our model in order to adapt the learners' differences?

Before introducing the tool implements our domain model, the following section describes our approach of domain model based on fuzzy prerequisite relationship.

### III. APPROACH OF DOMAIN MODEL BASED ON FUZZY PREREQUISITE RELATIONSHIP

In this approach, we use a domain model (learning hierarchy) predefined by one or more experts of a specific field. For that we determines in first phase an initial predefined domain model, in second phase we measure the variation of grades of learners, after that we transformed the data by using the fuzzification technique, then in the next phase we mine the association rules between the concepts. In the last two phases we propose to build the final model.

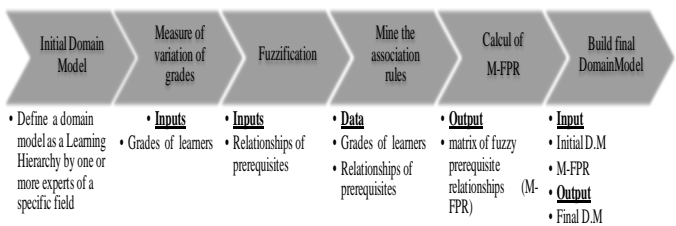

Fig. 1.  *Phases of approach* based on fuzzy prerequisite relationship

### IV. A TOOL IMPLEMENTS THE DOMAIN MODEL BASED ON FUZZY PREREQUISITE RELATIONSHIP

*A. Membership functions of fuzzy prerequisite relationship*

CPR a fuzzy subset of prerequisite relationships that can be classified as a correct prerequisite relationships between concept « i » and concept « j ».

$$CPR = \{(k, \mu_{CPR}(k))/k \in X\}$$

Where:

$\mu_{CPR}(k)$ Is the membership function of CPR, the values of this function present the relevance degree of each link « k » in the fuzzy set CPR.

RPR a fuzzy subset of links that can be classified as wrong prerequisite relationships between concept « i » and concept « j », but can be classified also as a correct prerequisite relationships between concept « j » and concept « i ».

$$RPR = \{(k, \mu_{RPR}(k))/k \in X\}$$

$\mu_{RPR}(k)$ is the membership function of RPR, the values of this function present the relevance degree of each link « k » in the fuzzy set RPR.

The definition of the two membership functions of fuzzy sets $\mu_{CPR}(k)$ and $\mu_{RPR}(k)$ is based on the indicator expressed as « variation of grades of all prerequisite relationships of initial domain model (ΔGrades) »

Where:

$$\mu_{CPR}(k) = \begin{cases} 0 & \text{if } \Delta Grades < S1 \\ \frac{-1}{S1}\Delta Grades + 1 & \text{if } S1 \leq \Delta Grades \leq 0 \\ \frac{-1}{S2}\Delta Grades + 1 & \text{if } 0 < \Delta grades \leq S2 \\ 0 & \text{if } \Delta Grades > S2 \end{cases}$$

$$\mu_{RPR}(k) = \begin{cases} 0 & \text{if } \Delta Notes < 0 \\ \frac{1}{S2}\Delta Notes & \text{if } 0 \leq \Delta Notes \leq S2 \\ \frac{-(\Delta Notes + S3)}{S3 - S2} & \text{if } S2 < \Delta Notes \leq S3 \\ 0 & \text{if } \Delta Notes > S3 \end{cases}$$

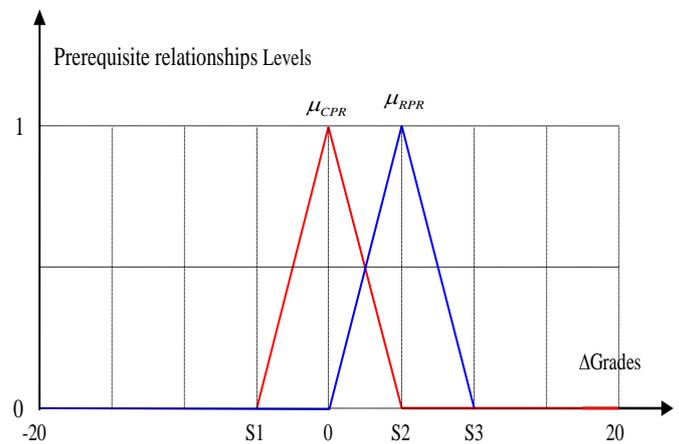

Fig. 2.  membership functions

*B. Description of the Tool*

The suggested tool is presented as a client-server application where experts in a specific field (Teachers of course) enable:

- Creating a new course (initial domain model) or adding a new concept to the existing courses.
- Adding the learners' grades for each concepts of initial domain model.
- Choose the three thresholds S1, S2 and S3 of the domain model based on fuzzy prerequisite relationship.
- Choose the threshold minimum of prerequisite relationships $\alpha_k$, this threshold indicates the prerequisite relationships meaningful in the domain model.
- Generation the final domain model

Our tool is an implementation of the course of java programming language.

*1. Creating an initial domain model of the course of Java programming language*

For this course were selected following 12 concepts:
1) Elementary of Java
2) Objects and Classes
3) Packages
4) Inner Classes
5) Flux I/O
6) Exceptions
7) Inheritance
8) Serialization
9) Interfaces
10) Polymorphism
11) Threads
12) Collections

Figure below shows the initial domain model of the course of Java programming language:

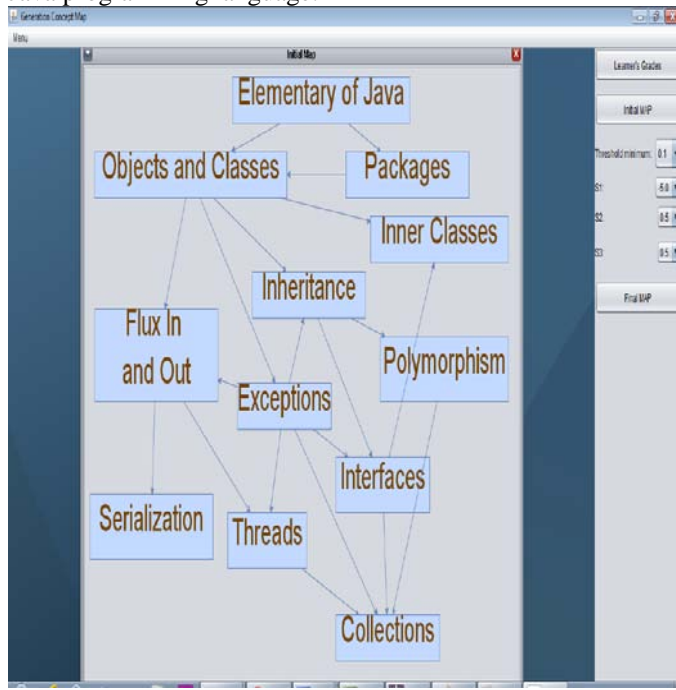

Fig. 3. Initial domain model of the course of Java

*2. Adding the learners' grades*

Figure below shows scores of 48 students obtained in the 12 concepts of the course of Java programming language:

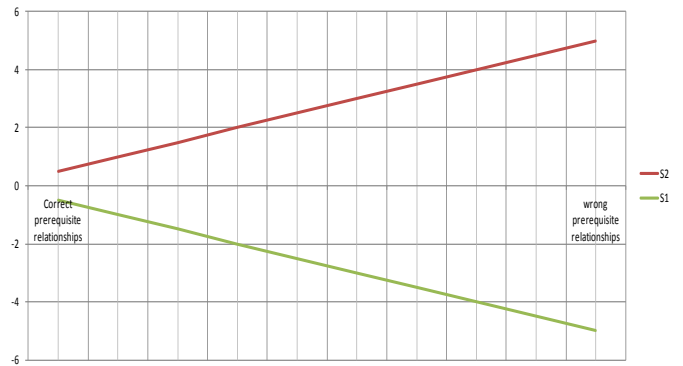

Fig. 4. learners' grades

*3. Choose the three thresholds S1, S2 and S3*

The three thresholds S1, S2 and S3 are defined in collaboration with experts in the field studied.

Based on our experience feedback on course of Java programming language the values the threshold of are chosen as follows:

Fig. 5. Possible values of threshold S1 & S2

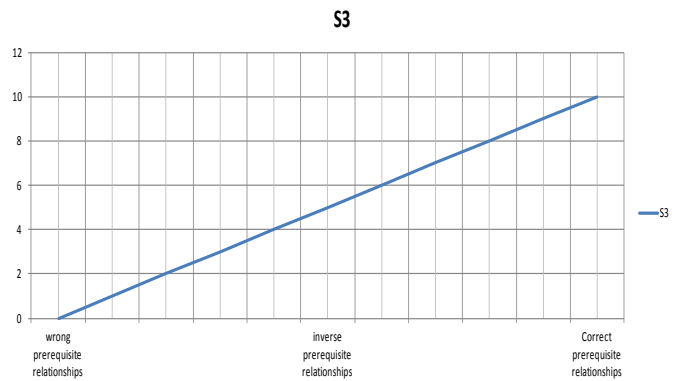

Fig. 6. Possible values of threshold S3

The combination of S1, S2 and S3 show that:

If the absolute values of S1 and S2 tends to 5 points the detection of correct prerequisites relationships are optimal, and if the value of S3 tends to 10 points detection of inverse prerequisites relationship is optimal.

### 4. Choose the threshold minimum of prerequisite relationships

The threshold minimum of prerequisite relationships $\alpha_k$, indicates the prerequisite relationships meaningful in the domain model.

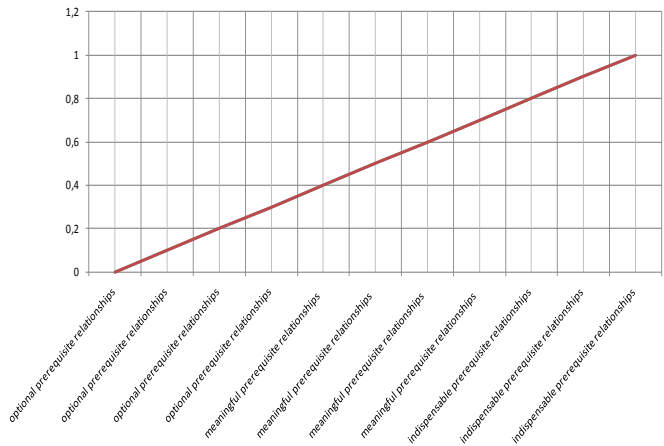

Fig. 7. Possible values of the threshold minimum of prerequisite relationship $\alpha_k$

The possible values of minimum of prerequisite relationships $\alpha_k$ are in range: $0 \leq \alpha_k \leq 1$:

For example, if we want select the indispensable prerequisite relationship in initial domain model we chose $0{,}8 \leq \alpha_k \leq 1$.

### 5. Generation the final domain model

a. Final domain model with:
$\alpha_k = 0{,}5$
S1 = variation of -5 grades
S2 = variation of 5   grades
S3 = variation of 10 grades

Then the two functions $\mu_{CPR}(k)$ and $\mu_{RPR}(k)$ becomes:

$$\mu_{CPR}(k) = \begin{cases} 0 & \text{if } \Delta Grades < -5 \\ \frac{1}{5}\Delta Grades + 1 & \text{if } -5 \leq \Delta Grades \leq 0 \\ \frac{-1}{5}\Delta Grades + 1 & \text{if } 0 < \Delta grades \leq 5 \\ 0 & \text{if } \Delta Grades > 5 \end{cases}$$

$$\mu_{RPR}(k) = \begin{cases} 0 & \text{if } \Delta Grades < 0 \\ \frac{1}{5}\Delta Grades & \text{if } 0 \leq \Delta Grades \leq 5 \\ \frac{-1}{5}\Delta Grades + 2 & \text{if } 5 < \Delta grades \leq 10 \\ 0 & \text{if } \Delta Grades > 10 \end{cases}$$

And the final domain model of the course of Java programming language is:

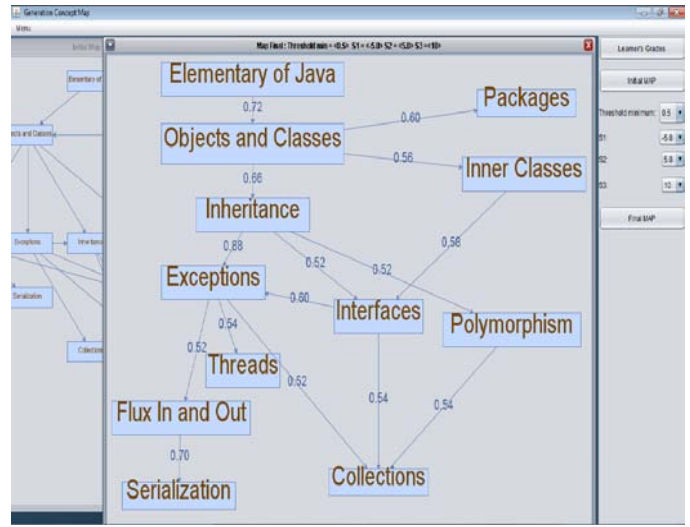

Fig. 8. *Final* domain model of the course of Java programming language (1)

These values of the parameters give the best results in practice.

b. Final domain model with:
$\alpha_k = 0{,}2$
S1 = variation of -5 grades
S2 = variation of 5   grades
S3 = variation of 10 grades

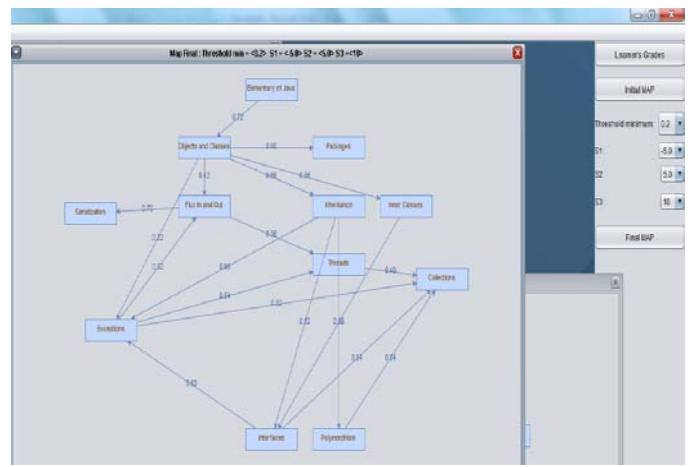

Fig. 9. *Final* domain model of the course of Java programming language (2)

These parameter values have selected non-significant relationship.

c. Final domain model with:
$\alpha_k = 0{,}3$
S1 = variation of -3 grades
S2 = variation of 3 grades
S3 = variation of 5 grades

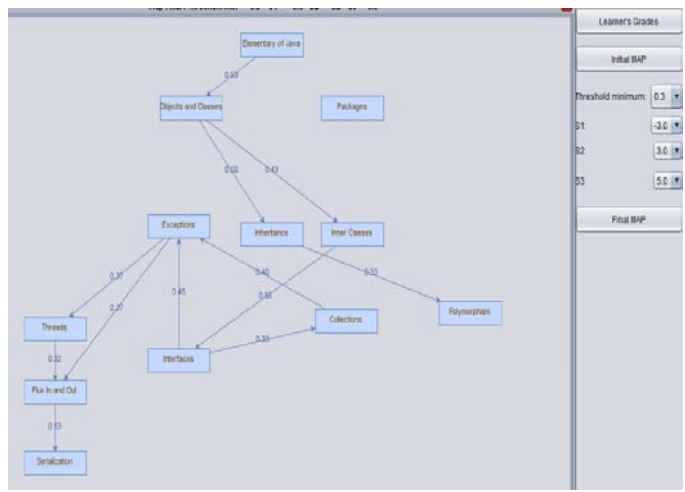

Fig. 10. *Final* domain model of the course of Java programming language (3)

These parameter values have deleted some important relationships.

## V. CONCLUSION

In this paper we present an implementation of a practical application of fuzzy logic techniques to the domain model of a specific field.

This tool allow expert in a specific field to creating an initial domain model, choose the value of the indicator of the prerequisite relationships meaningful in the domain model.
And also allow them to configure the three thresholds of the final domain model.
Based on experimental the values of the parameters that give the best adaption of the learners' differences on the course of Java programming language are:

$\alpha_k = 0{,}5$
S1 = variation of -5 grades
S2 = variation of 5 grades
S3 = variation of 10 grades.

REFERENCES

[1] Aajli and Afdel, "A computer adaptive assessment system for E-Learning and E-Recruitment based on a new measuring skills approach International", Journal of Educational Technology Letters Volume 3, Number 1, June, 2013
P. 42-51, Online: http://www.ier-institute.org/2163_4246.html

[2] Aajli and Afdel, "A New Approach of Learning Hierarchy Construction Based on Fuzzy Logic" Int. Journal of Engineering Research and Applications, Vol. 4, Issue 10( Part - 3), 2014, pp.58-66

[3] Alkhazaleh and Salleh, "Fuzzy Soft Multiset Theory, Abstract and Applied Analysis", 2012, article ID 350600, 20 p.

[4] Al-Sarem, Bellafkih and Ramdani "Mining Concepts' Relationship Based on Numeric Grades", JCSI International Journal of Computer Science Issues, Vol. 8, Issue 4, No 2, July 2011

[5] Ana J. Viamonte, The Computer in the Mathematics Teaching, WSEAS Transactions on Advances in Engineering Education, 2010, Vol. 7, No. 3, pp.63-72.

[6] Anohina, A., & Grundspenkis, J. (2009, June). Scoring concept maps: an overview. In Proceedings of the International Conference on Computer Systems and Technologies and Workshop for PhD Students in Computing P.78.

[7] Berlin Heidelberg. Crowder, N. A. Intrinsic and extrinsic programming. Programmed Learning and Computer-Based Instruction. New York: John Wffley, (1962).

[8] Brusilovsky, P. (1996). Adaptive Hypermedia: An Attempt to Analyze and Generalize. Proceedings of First International Conference on Multimedia, Hypermedia and Virtual Reality 1994.

[9] Brusilovsky, P. (2000) "Adaptive Hypermedia: From Intelligent Tutoring Systems to Web-Based Education", in Proc. Intelligent Tutoring Systems, pp.1-7.

[10] Brusilovsky, P., (1998). Methods and Techniques of Adaptive Hypermedia. Adaptive Hypertext and Hypermedia. pp.1-43.

[11] Cañas, A. J et al. "Concept maps", Integrating knowledge and information visualization. In Knowledge and information visualization, Springer, 2005 (pp. 205-219).

[12] Das and Martins, "A survey on automatic text summarization",2007. Online:http://www.cs.cmu.edu/_afm/Home_files/Das_Martins_survey_summarization.pdf

[13] Danis, Schubauer-Leoni and Weil-Barais, "Interaction, Acquisition de connaissances et Développement", Bulletin de Psychologie, 2003.

[14] D. Dagger, O. Conlan and Vincent P. Wade, "an architecture for candidacy in adaptive e-learning systems to facilitate the reuse of learning resources". In World Conference on E-Learning in Corporate, Government, Healthcare and Higher Education E-Learn, 2003, pp. 49–56.

[15] De Cock, Bodenhofer, and Kerre, "Modelling Linguistic Expressions Using Fuzzy Relations", Proceedings 6th International Conference on Soft Computing, Iizuka, Japan, 1-4 october 2000, p. 353-360.

[16] Ertmer, P. A., & Newby, T. J. Behaviorism, cognitivism, constructivism: Comparing critical features from an instructional design perspective. Performance improvement quarterly, (1993) , P. 50-72.

[17] Gagne, R. M, "Learning hierarchies", Educational psychologist, 1968, P. 1-9.

[18] P. De Bra, N. Stash and D. Smits. "Creating adaptive web-based applications," In Tutorial at the 10th International Conference on User Modeling, 2005, Edinburgh, Scotland.

[19] P. Toth, "Online learning behavior and web usage mining," WSEAS Transactions on Advances in Engineering Education, vol. 10, no. 2, 2013, pp. 71-81.

[20] R. Rahamat, P. M. Shah, R. Din, S. N. Puteh, J.A. Aziz, H. Norman, M. A.Embi, "Measuring learners' perceived satisfaction towards e-learning materials and environment," WSEAS Transactions on Advances in Engineering Education, 2012, vol. 9, no. 3, pp. 72-83.

[21] Skinner, B. F, "Programmed Instruction Revisited". Phi Delta Kappan, 1986, 103.

[22] Zadeh, "The concept of a linguistic variable and its application to approximate reasoning–I", Inform. Sci., v. 8, pp. 199–249, 1975.

[23] Zubrinic et al. "Automatic creation of a concept map" International Journal Expert Systems with Applications, Volume 39 Issue 16, November, 2012